\documentclass[traditabstract]{./aa}
\usepackage{amsmath, amssymb, upgreek}

\usepackage{graphicx}
\usepackage{txfonts}
\usepackage{natbib}

\def\vv{\varv}
\newcommand{\added}{\textbf}
\def\mpsec{m$\cdot$s$^{-1}$}
\def\phidip{$\Phi_{\rm dipole}$}
\def\rmax{$R_{\rm max}$}

\begin{document}
   \title{Are the strengths of solar cycles determined by 
          converging flows towards the activity belts?}


   \author{R.H.~Cameron\inst{\ }
          \and M.~Sch\"ussler\inst{\ }}

   \institute{Max-Planck-Institut f\"ur Sonnensystemforschung,
              Max-Planck-Str. 2, 37191 Katlenburg-Lindau, Germany\\
              \email{cameron@mps.mpg.de, schuessler@mps.mpg.de}
                       }
   \date{Received ; accepted }

\abstract{It is proposed that the observed near-surface inflows towards
  the active regions and sunspot zones provide a nonlinear feedback
  mechanism that limits the amplitude of a Babcock-Leighton-type solar
  dynamo and determines the variation of the cycle strength. This
  hypothesis is tested with surface flux transport simulations including
  converging latitudinal flows that depend on the surface distribution
  of magnetic flux. The inflows modulate the build-up of polar fields
  (represented by the axial dipole) by reducing the tilt angles of
  bipolar magnetic regions and by affecting the cross-equator transport
  of leading-polarity magnetic flux. With flux input derived from the
  observed record of sunspot groups, the simulations cover the period
  between 1874 and 1980 (corresponding to solar cycles 11 to 20). The
  inclusion of the inflows leads to a strong correlation of the
  simulated axial dipole strength during activity minimum with the
  observed amplitude of the subsequent cycle.  This in agreement with
  empirical correlations and in line with what is expected from a
  Babcock-Leighton-type dynamo. The results provide evidence that the
  latitudinal inflows are a key ingredient in determining the amplitude
  of solar cycles.}
   
   \keywords{Sun: dynamo -- Sun: activity -- Sun: surface magnetism}
   \authorrunning{Cameron \& Sch\"ussler}
   \titlerunning{Strengths of solar cycles}
   \maketitle
%

\section{Introduction}

The correlation between the amplitude of the solar polar field around
activity minimum and the strength of the subsequent cycle
\citep[e.g.,][see, however, \citeauthor{Layden:1991},
\citeyear{Layden:1991}]{Schatten:etal:1978, Choudhuri:2008, Cameron10}
suggests that the Sun's axial dipole field as observed around activity
minimum reflects the source for the generation of the toroidal magnetic
flux of the subsequent cycle, as opposed to being a mere epiphenomenon
of an otherwise fully hidden subsurface dynamo process
\citep{Cameron:Schuessler:2007, Schuessler:2007}.  Such a role of the
polar field is a key feature of the Babcock-Leighton (BL) dynamo model
\citep[see][for a review of solar dynamo theory]{Charbonneau10}.  In
this type of dynamo, the poloidal magnetic field results from the
systematic tilt (with respect to the azimuthal/longitudinal direction)
of sunspot groups and bipolar magnetic regions in combination with the
subsequent redistribution of their magnetic fluxes by near-surface
flows. The crucial mechanism for the reversal of the Sun's
global dipole field is the preferred transport of preceding-polarity
flux of bipolar magnetic regions across the equator into the other solar
hemisphere \citep{Cameron:Schuessler:2007}, which leads to the reversal
of the polar fields and the build-up of an axial dipole field of
opposite polarity.  Toroidal flux is then generated by the action of
differential rotation winding up poloidal field lines threading the Sun.

So far we have described the BL dynamo as a linear process. In addition,
at least one nonlinear feedback mechanism is required that limits the
amplitude of the generated magnetic field and controls the variation of
the cycle amplitude. In the framework of the BL dynamo, potential
feedback mechanisms include the back-reaction of the magnetic field on
1) differential rotation, 2) the tilt angles of sunspot groups, and 3)
cross-equator transport of magnetic flux.

Concerning possibility 1), the only empirically well-established
cycle-related variation of the differential rotation are zonal flows of
a few \mpsec, which at low latitudes are associated with the activity
belts \citep{Howard:Labonte:1980, Zhao04, Gizon04,
Gizon:Birch:2005}. Clear evidence for a more global reaction on
differential rotation that would affect the generation of the toroidal
magnetic field has not been obtained so far.

As to mechanism 2), the back-reaction on the tilt angles of sunspot
groups, \citet{Dasi-Espuig10} found cycle-to-cycle variations of the
average tilt angle, which are anti-correlated with the cycle strength:
strong cycles show smaller average tilt angles than weak cycles. Since
the tilt angles affect the strength of the source for the poloidal
field in the BL dynamo, their anticorrelation with cycle strength
provides a possible nonlinear saturation mechanism.  \cite{Cameron10}
included the observed tilt-angle variations in a surface flux
transport (SFT) simulation with sources taken from the combined Royal
Greenwich Observatory (RGO) and the Solar Optical Observing Network
(SOON) sunspot records.  Considering the period for which
tilt-angle data are available (1913--1986), they found that the
maximum of the polar field (occurring around activity minimum of a
given cycle) in the SFT simulation is correlated with the observed
strength of the subsequent activity cycle. This suggests that the
anti-correlation of the average tilt angle with cycle strength
contributes to the variations of the cycle amplitudes and thus to the
nonlinear feedback by which the solar dynamo saturates at the observed
levels.

Which physical process could lead to the observed anti-correlation?  One
possibility is a change in the strength or effectiveness of the Coriolis
force that tilts magnetic flux tubes rising in the convection zone
\citep[e.g.,][]{Dsilva:Choudhuri:1993, Fan:etal:1994,
Caligari:etal:1995,Caligari:etal:1998,Fan:2009}. In the absence of
sufficient observational information about the subsurface structure and
dynamics of the magnetic field, it is difficult to evaluate this
mechanism. Another possible process is more amenable to observation and
quantitative test: tilt-angle changes caused by the redistribution of
magnetic flux after emergence by near-surface flows resulting from the
presence of the magnetic field. A well-observed example of such flows
are the inflows towards active regions and the activity belts
\citep[e.g.,][]{Haber02, Zhao04, Gizon04, Gizon08, Svanda08,
Gonzalez10}, possibly resulting from a temperature deficit due to the
excess radiance of small-scale magnetic flux concentrations
\citep{Spruit03, Gizon08}.  Collectively, these inflows drive mean flows
towards the activity belts and zonal flows, whose amplitudes depend on
the amount of magnetic flux present in active regions, i.e., on the
overall strength of a cycle.  The transport of active-region flux by
both the converging latitudinal flows and the longitudinal zonal flows
tends to reduce the tilt angles of the corresponding bipolar magnetic
regions.  This mechanism was recently studied by \cite{Jiang10b} with
SFT simulations including an ad-hoc model of the inflows in terms of
latitudinal flows towards the activity belts.  Considering individual
bipolar regions as well as synthetic solar cycles, they found that the
inflows reduce the lateral separation of the magnetic polarities,
leading to a weakening of the contribution of a given bipolar region
to the polar field.  Building on these results,
\citet{Jiang:etal:2011a,Jiang:etal:2011b} included cycle-dependent tilt
angles in their reconstruction of the evolution of the surface flux and
open heliospheric flux since 1700.

There is another aspect of the converging inflows towards active
regions, which is related to possibility 3) for a nonlinear effect
listed further above: a feedback on the cross-equator transport of
magnetic flux.  For active regions not too far from the equator, the
inflows often extend over the equator \citep[see, e.g., Fig.~18
in][]{Gizon:etal:2010}, which enhances the otherwise purely diffusive
cross-equator transport of preceding-polarity flux and, therefore, tends
to increase the resulting polar field. This effect is weaker for strong
cycles because then active regions start to emerge at higher latitudes
and the activity maximum is reached earlier (the Waldmeier effect), at a
time when the activity belts are farther away from the
equator. Therefore, this represents a negative feedback on the growth of
the polar field.

In this paper, we present results of SFT simulations including mean
latitudinal flows towards the activity belts, which were determined
self-consistently from the simulated distribution of magnetic flux, thus
introducing a nonlinear element into the SFT model. Using the RGO/SOON
sunspot group data to determine the flux in emerging bipolar magnetic
regions, we followed the evolution of the surface flux from 1874 until
1980. In order to evaluate the possible relevance of the nonlinearity
for determining the cycle strength in the framework of a BL dynamo, we
considered the correlation between the maximum of the magnetic flux
associated with the axial dipole component in a given cycle resulting
from the SFT simulation with the observed amplitude of the subsequent
cycle.  

The advantage of the approach chosen here is that it deals exclusively
with observationally constrained near-surface processes and thus does
not require a complete dynamo model in order to evaluate the potential
of the nonlinearity. While the surface evolution addressed by SFT
simulations is constrained by observations, a full dynamo model involves
assumptions of the subsurface dynamics, which at present are essentially
unconstrained.

\section{Surface flux transport simulation}
\label{sec:sftm}

\subsection{Transport equation}
\label{subsec:equation}

We used the SFT code of \citet{Baumann:etal:2004} in the version
described by \citet{Jiang10b} to simulate the evolution of the magnetic
flux at the solar surface under the influence of large-scale surface
flows (differential rotation and meridional flow) and a diffusivity
representing the random motion of the magnetic flux elements due to the
(numerically unresolved) supergranular flows. The magnetic field at the
solar surface is assumed to be radial. The governing equation of the SFT
model is
\begin{eqnarray}
\frac{\partial B}{\partial t}=
     &-&\omega(\lambda) \frac{\partial B}{\partial \phi} 
      -\frac{1}{R_{\odot}\cos \lambda} \frac{\partial}{\partial \lambda} 
     \bigg\{\Big[\vv(\lambda)+\Delta \vv(\lambda,t)\Big]
           B \cos\lambda \bigg\} \nonumber \\ \noalign{\vskip 2mm} 
     &+& \eta_H\left[\frac{1}{R_{\odot}\cos \lambda} 
     \frac{\partial}{\partial \lambda} 
     \left(\cos \lambda \frac{\partial B}{\partial \lambda} \right)
         + \frac{1}{R_{\odot}^2\cos^2 \lambda}\frac{\partial^2
     B}{\partial \phi^2}\right] \nonumber \\ \noalign{\vskip 2mm} 
    &+& S(\lambda,\phi,t)+D(\eta_r)\,,     
\label{eqn:sft}
\end{eqnarray}
where $B(\lambda,\phi)$ denotes the radial component of the magnetic
field, $\lambda$ is the heliographic latitude, and $\phi$ is the
heliographic longitude. The expressions for the synodic differential
rotation \citep[from][]{Snodgrass83},
\begin{eqnarray}
\omega(\lambda)=13.38-2.30 \sin^2 \lambda-1.62 \sin^4 \lambda 
       \; [^{\circ}/\mathrm{day}],
\end{eqnarray}
and for the time-independent part of the meridional flow
\citep[cf.][]{van_Ballegooijen98},
\begin{equation}
\vv(\lambda)=
\begin{cases} 11\sin(\pi\lambda/75^{\circ})\; \text{m}\,\text{s}^{-1}& 
       \text{where} \mid \lambda \mid \le 75^{\circ}\\
       0 & \text{otherwise},
\end{cases}
\label{eqn:mer}
\end{equation}
as well as the choice of the horizontal and radial diffusivities
\citep{Baumann:etal:2006}, $\eta_H=250$~km$^{2}$s$^{-1}$ and $D=0$, were
the same as used in \cite{Cameron10}.  The velocity
$\Delta\vv(\lambda,t)$ in Eq.~(\ref{eqn:sft}) represents the
longitude-averaged inflows toward the activity belts, which are
described by a magnetic-field-dependent local perturbation of the
meridional flow; its construction is explained in
Sec.~\ref{subsec:inflows}. This term renders the model nonlinear.

\begin{figure*}[ht!]
   \flushleft
   \includegraphics[scale=1.]{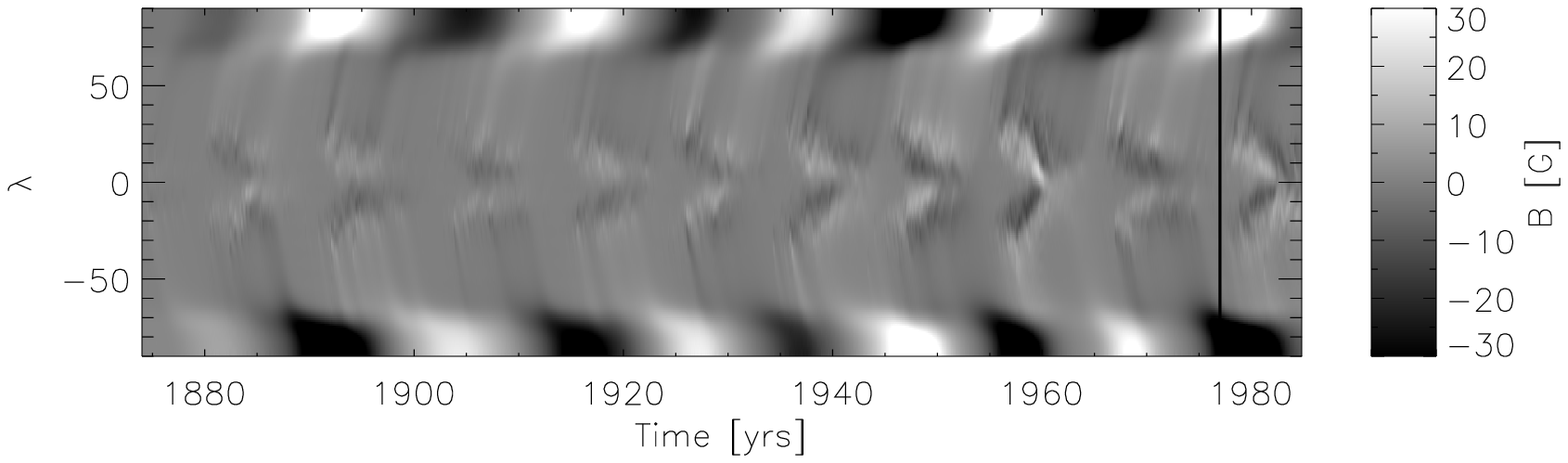}
   \vskip 3mm
   \includegraphics[scale=1.]{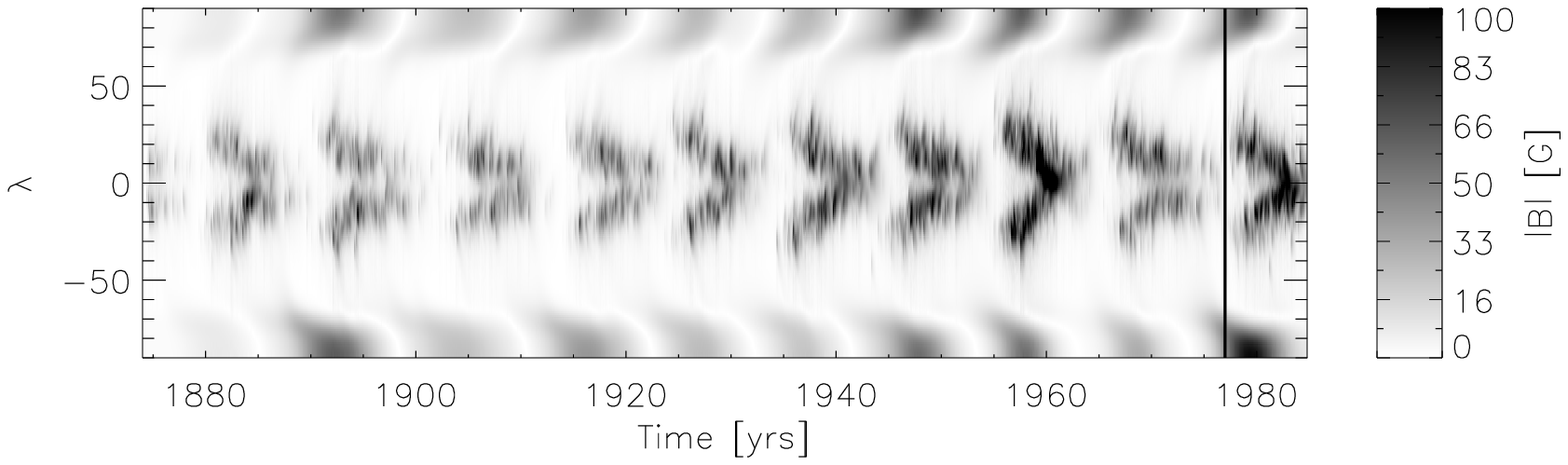}
   \vskip 3mm
   \includegraphics[scale=1.]{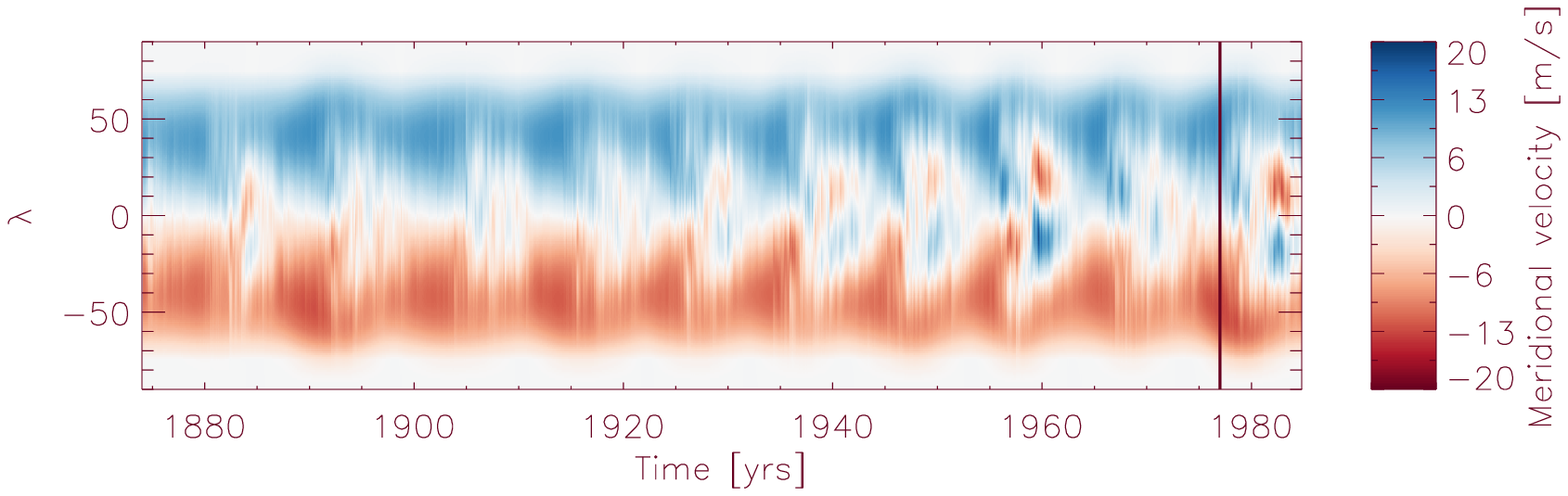}
   \vskip 3mm
   \includegraphics[scale=1.]{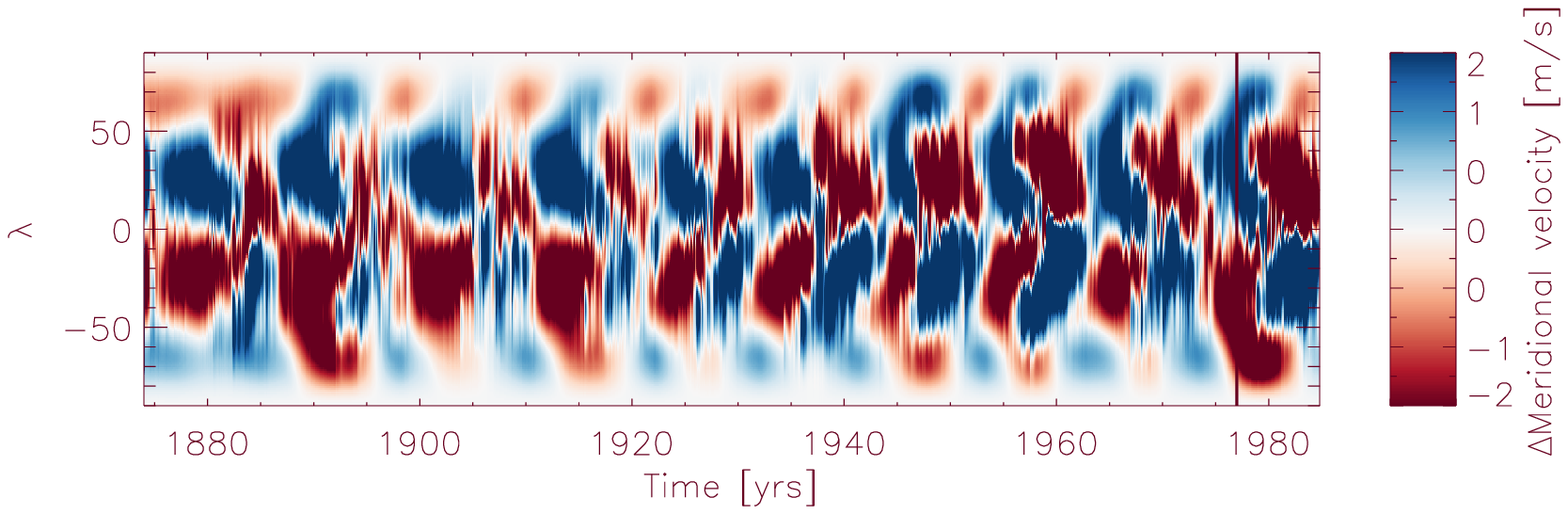}
   \caption{Time-latitude diagrams of various quantities from a SFT
            simulation run with magnetic-field dependent meridional
            inflows towards the activity belts. From top to bottom:
            longitudinally averaged signed radial magnetic field,
            longitudinally averaged unsigned radial magnetic field,
            meridional flow velocity, and meridional flow perturbation
            ($\Delta\vv(\lambda,t)$, saturated for values exceeding
            $\pm2\,$\mpsec). Black vertical lines indicate the
            termination of the RGO sunspot data in 1976.}
         \label{fig:time_latitude}                                     
\end{figure*}

\subsection{Flux emergence}
\label{subsec:fluxem}

The source term, $S(\lambda,\phi,t)$, in Eq.~(\ref{eqn:sft}) represents
the emergence of magnetic flux at the surface in the form of bipolar
magnetic regions appearing at the latitudes and longitudes of sunspot
groups taken from the combined RGO/SOON sunspot record
\citep{Balmaceda:etal:2009} at the time of their maximum surface
area. The sunspot group size was converted to magnetic flux flux as
described in \citeauthor{Cameron10} (\citeyear{Cameron10}, and
references therein), the calibration being based upon the total unsigned
magnetic flux derived from the synoptic magnetic maps%
\footnote{{\tt http://soi.stanford.edu/magnetic/index6.html}} %
taken with the Michelson Doppler Imager (MDI) instrument onboard of
the Solar Heliospheric Observatory (SoHO) spacecraft. We assumed that
the latitude dependence of the initial tilt angle, $\alpha$, of the
new bipolar magnetic regions has the form $\alpha= T \sqrt{\lambda}$,
using $T=1.42$ for all cycles (angles measured in degrees), so that
cycle-dependent changes of the tilt angles result solely from the
magnetic-field dependent inflows toward the activity belts.

The transition from the RGO data covering the period up to 1976 to the
subsequent SOON data involves cross-calibration issues.  For instance,
the minimum area required for a sunspot group to be included changed
from 1 millionth of a hemisphere in the RGO data to 10 millionths of a
hemisphere in the SOON data. In order to account for this change,
\citet{Balmaceda:etal:2009} used other datasets overlapping both the RGO
and SOON data to determine a correction factor of approximately 1.4 for
the sunspot group areas in the SOON data \citep[see also
][]{Hathaway05}.  While a global correction factor might be adequate for
the total sunspot areas, applying it uniformly to individual sunspot
groups of all sizes (as used in the source term of SFT simulations) is
hardly justified since it probably grossly overestimates the areas of
the large sunspot groups. This affects particularly strongly the
variation of the polar field, to which the large sunspot groups
contribute most because of their large latitude separations and
fluxes. Since the overestimated sources lead to unrealistic
results for the polar fields (and the closely related axial dipole
component), which are at our focus of interest in this paper, we
consider the SFT results only until the activity minimum between cycles
20 and 21, which nearly coincides with the termination of the RGO data
in 1976.

\subsection{Latitudinal inflows}
\label{subsec:inflows}

We treated the cumulative effect of the inflows toward active regions in
terms of a local perturbation, $\Delta \vv(\lambda,t)$, of the
axisymmetric meridional flow. The amplitude of the background meridional
flow was kept constant. This is consistent with the result of
\citet{Cameron10b}, who showed that the active-region inflows fully
explain the variations in the low-degree spherical-harmonic
decomposition of the observed meridional flow \citep{Hathaway10},
without requiring a change in the overall meridional flow amplitude. A
time-independent global meridional circulation is also consistent with
the finding by \cite{Hathaway:2011} that the latitudinal drift rates of
the activity belts are the same for all cycles when measured with respect
to the epochs of the cycle minima. The perturbation of the meridional
flow in our SFT simulations was therefore modeled using almost the same
procedure as in \cite{Cameron10b}. Taking the instantaneous latitude
profile of the azimuthally averaged unsigned radial magnetic surface
field,
\begin{equation}
\langle |B| \rangle (\lambda,t)=  \frac{1}{2\pi} \int_{0}^{2\pi} 
        |B(\phi,\lambda,t))| \, {\mathrm{d}}\phi\,,
\end{equation}
we set the inflow speed towards the activity belts proportional to the
smoothed derivative of $\langle |B| \rangle$ with respect to latitude,
viz.
\begin{equation}
\Delta \vv(\lambda,t)=c_0  \int 
   \left( \frac{\cos\lambda'}{\cos30^\circ} \right)
   \frac{ d \langle |B| \rangle}{{\mathrm{d}}\lambda'} 
   {\mathrm{e}}^{-(\lambda-\lambda')^2/\sigma} 
   {\mathrm{d}}\lambda'\\,
\label{eq:delta_v}
\end{equation}
where the choice of $\sigma$ effects a smoothing in latitude with a full
width at half maximum of 20$^{\circ}$. The prefactor
$c_0=9.2$~m\,s$^{-1}$G$^{-1}$deg was calibrated in \cite{Cameron10b} by
requiring that the amplitude of the inflow should be comparable to that
reported by \cite{Gizon:etal:2010} from helioseismic observations during
cycle 23. The only difference to the procedure of \cite{Cameron10b} is
the factor ($\cos\lambda/\cos30^\circ$) in Eq.~(\ref{eq:delta_v}), which
suppresses unrealistically strong flow perturbations that would
otherwise result at high latitudes from the gradient of the polar
fields.

For a nonlinear SFT simulation including the effect of
magnetic-field-dependent inflows, Fig.~\ref{fig:time_latitude} shows
time-latitude diagrams of the longitude-averaged radial magnetic field
(signed and unsigned) as well as of the meridional flow speed and its
perturbations, $\Delta \vv$, due to the modeled inflows.  Stronger
cycles are associated with bigger perturbations of the meridional
flow. For very high cycle amplitude, the flow direction may even locally
reverse, which can be seen most prominently during cycle 19. Depending
on the asymmetry between the magnetic field distributions in the two
hemispheres, cross-equator flows occasionally arise.

\begin{figure}[ht!]
   \flushleft
   \includegraphics[width=\linewidth]{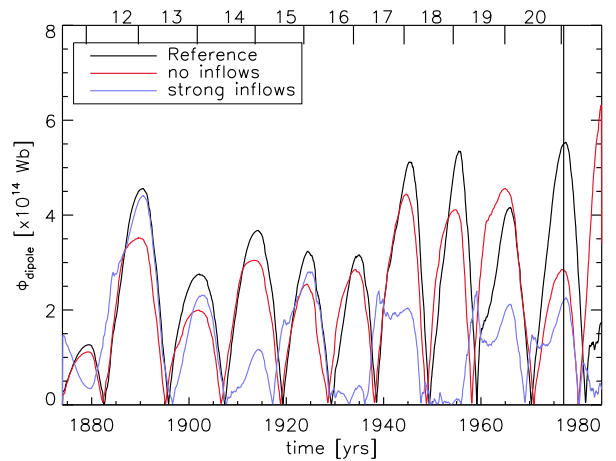}   
      \caption{Magnetic flux corresponding to the axial dipole,
               $\Phi_{\rm dipole}$, from the SFT simulation with inflows
               (reference case, black curve), without inflows (red
               curve), and with strong inflows (amplitude multiplied by
               1.5 with respect to the reference case, blue curve). The
               black vertical line indicates the termination of the RGO
               sunspot data in 1976.}
         \label{fig:dipolar}
\end{figure}

\begin{figure}[h!]
   \flushleft
   \includegraphics[scale=0.5]{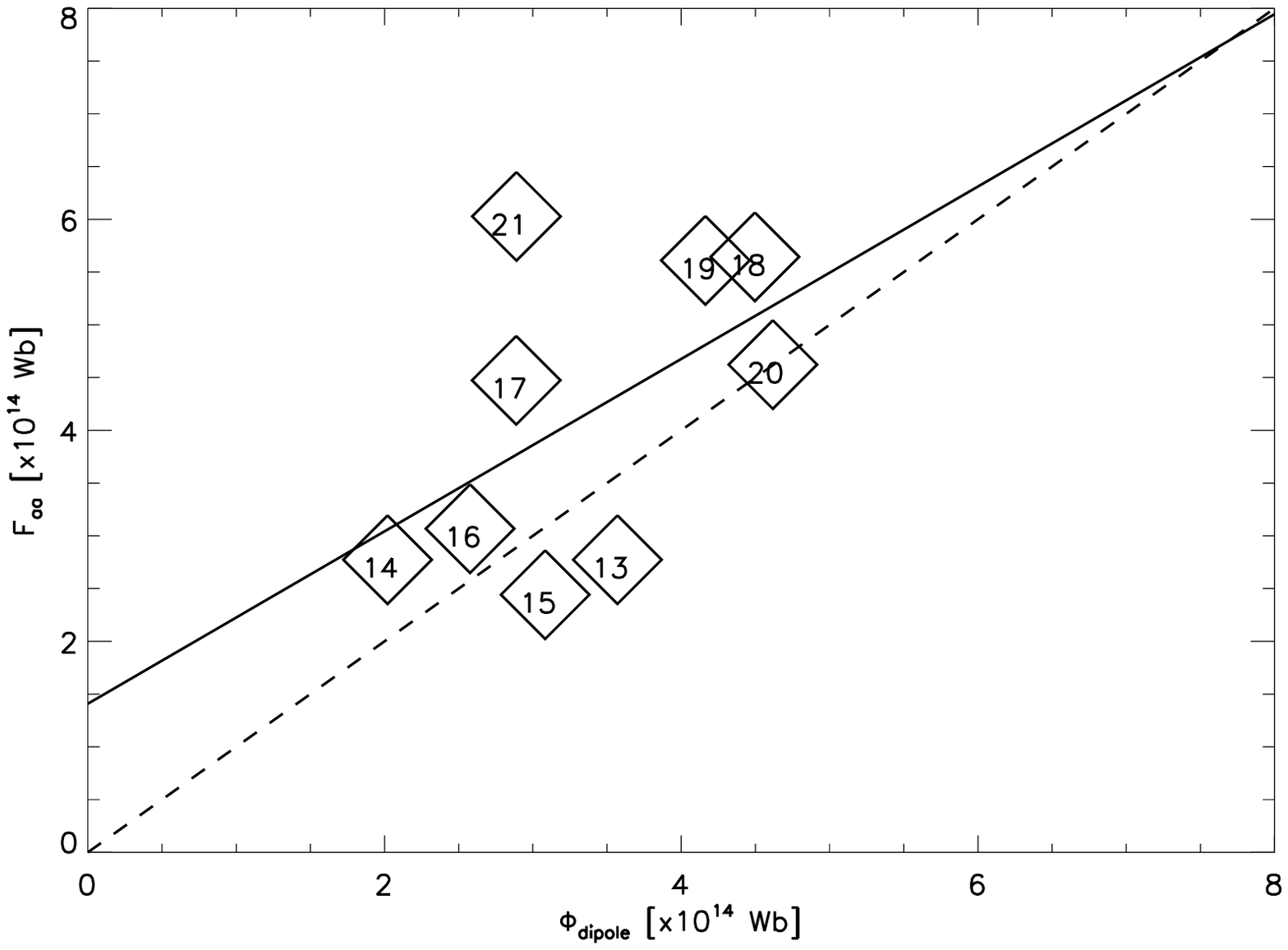}\\
   \includegraphics[scale=0.5]{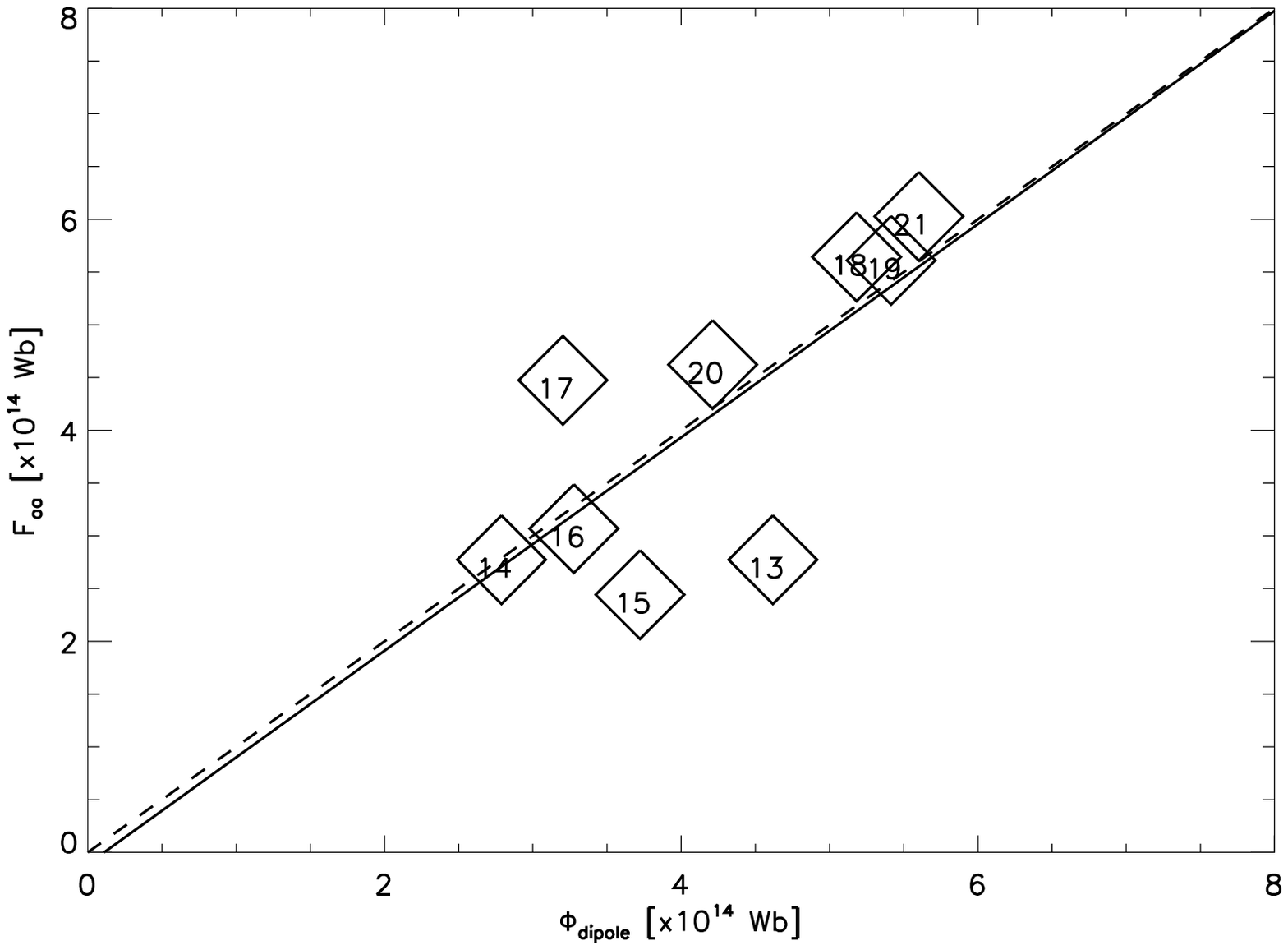}
      \caption{Maxima of $\Phi_{\rm dipole}$ from SFT simulations
               vs. open flux during activity minima as determined from
               the measured geomagnetic $aa$-index
               \citep{Lockwood:etal:2009}. {\em Upper panel:} case
               without inflows. {\em Lower panel:} reference case with
               inflows. Numbers within symbols give the index of the
               cycle following the respective activity minimum. Full
               lines indicate regression curves, dashed lines the
               bisectrix {\added{of the first quarter plane}}. }
         \label{fig:dipolar_vs_aa}
\end{figure}

\section{Results}
\label{subsec:results}

\subsection{Effects of the inflows on the axial dipole}
\label{sec:cycles}

The latitudinal inflows towards the activity belts affect the
near-surface evolution of the magnetic field in a nonlinear fashion,
since they depend on the magnetic field distribution and strength. To
investigate the potential of this effect as a nonlinear feedback in
flux-transport dynamos, we compare SFT simulations with and without
inflows. A relevant quantity for the amplitude of flux-transport dynamos
is the maximum axial dipole component of the surface field around
activity minimum. This field represents the main part of the poloidal
field from which the toroidal field of the next activity cycle is
generated by differential rotation. Note that the axial dipole component
is axisymmetric and therefore independent of the presence of zonal flows
since any differential rotation does not affect the longitude-averaged
radial surface field in a SFT simulation
\citep[e.g.,][]{Baumann:etal:2004}.

The time development of the magnetic flux associated with the axial
dipole, \phidip, for cycles 12 through 20 is shown in
Fig.~\ref{fig:dipolar}. Three SFT simulation runs are compared: the
`reference' case (cf. Fig.~\ref{fig:time_latitude}) with calibrated
inflows towards the activity belts (black line), the case without
inflows (red line), and a case for which the inflows were enhanced by a
factor 1.5 with respect to the reference case (blue line).  The
first cycle in the simulation is not considered further since it is
still dominated by the initial condition \citep[for details,
see][]{Cameron10}. In both the reference case and the case with no
inflows, \phidip\ peaks at or shortly before the activity minima.

For all cycles except cycles 19, we see that the inflows in the
reference case tend to increase \phidip. We can understand this result
by considering that the preferential cross-equator transport of
leading-polarity magnetic flux in tilted bipolar magnetic regions is the
essential effect that produces a flux imbalance in each hemisphere,
leading to the polarity reversal and build-up of the opposite-polarity
axial dipole field \citep[e.g.,][]{Cameron:Schuessler:2007}.  There are
two opposing effects of the latitudinal inflows on the cross-equator
transport: 1) for low-latitude active regions, the inflows are broad
enough to cross the equator and thus enhance the flux transport from the
other hemisphere, increasing the flux imbalance and thus eventually
strengthening the axial dipole, and 2) the convergence of the inflows
reduces the tilt angle, counteracting the diffusive latitudinal
spreading of a bipolar region, and thus weakening its eventual
contribution to the axial dipole. For weak to moderate cycles, the first
effect dominates and the enhanced cross-equator transport leads to
stronger axial dipole fields around activity minima. For stronger
cycles, the combination of the Waldmeier effect (whereby stronger cycles
peak earlier) and the fact that activity starts at higher latitudes
before propagating towards the equator results in the activity belts
being further away from the equator during the maximum phase of such
cycles. This suppresses the flux transport by cross-equatorial flows, so
that eventually the opposing effect 2) becomes dominant: the inflows act
against the diffusion to reduce the tilt angle and thus reduce the
diffusive cross-equatorial transport of magnetic
flux. Fig.~\ref{fig:dipolar} shows this clearly for the minimum periods
between cycles 19, 20, and 21: while the maxima of the axial dipole flux
in the case without inflows (red curve) largely follow the amplitudes of
the preceding cycles in a linear manner, including the inflows (black
curve) leads to a weakened axial dipole after cycle 19 and a much
stronger dipole after cycle 20.

We studied the sensitivity of the axial dipole on the strength of the
inflows by running SFT simulations with inflow amplitudes multiplied by
factors of 0.5 and 1.5, respectively, with respect to the reference
case. While in the case of weakened inflows the time evolution of
\phidip\ is essentially the same as in the case of no inflows (red curve
in Fig.~\ref{fig:dipolar}), for the enhanced inflows the deviations from
the reference case are drastic (blue curve in Fig.~\ref{fig:dipolar}).
The overly strong reduction of the cross-equator flux transport in this
case leads to some almost failed polar field reversals and a resulting
strong 22-year periodicity of \phidip.

For a comparison of the SFT models with empirical results,
Fig.~\ref{fig:dipolar_vs_aa} shows the relation between the maxima of
\phidip\ from our SFT simulations and the open heliospheric magnetic
flux during these periods determined from the variations of the
geomagnetic $aa$-index \citep{Lockwood:etal:2009}. We may safely assume
that the open magnetic flux during activity minimum periods is strongly
dominated by the axial dipole \citep[e.g.,][]{Wang:Sheeley:2009}. The
upper panel of Fig.~\ref{fig:dipolar_vs_aa} gives the result for the
simulation without inflows, which shows a weak $(r=0.55)$ and
insignificant $(p=0.15)$ correlation; there is also a clear deviation of
the regression line (solid) from the bisectrix (dashed). In contrast,
for the reference case with inflows (lower panel of
Fig.~\ref{fig:dipolar_vs_aa}) we find a good correlation $(r=0.74,
p=0.022)$ and a regression line nearly coinciding with the bisectrix.
It is obvious from the blue curve in Fig.~\ref{fig:dipolar} that the
case with enhanced inflows is inconsistent with the empirically
determined open fluxes.
 
\begin{figure*}[ht!]
   {\includegraphics[scale=0.5]{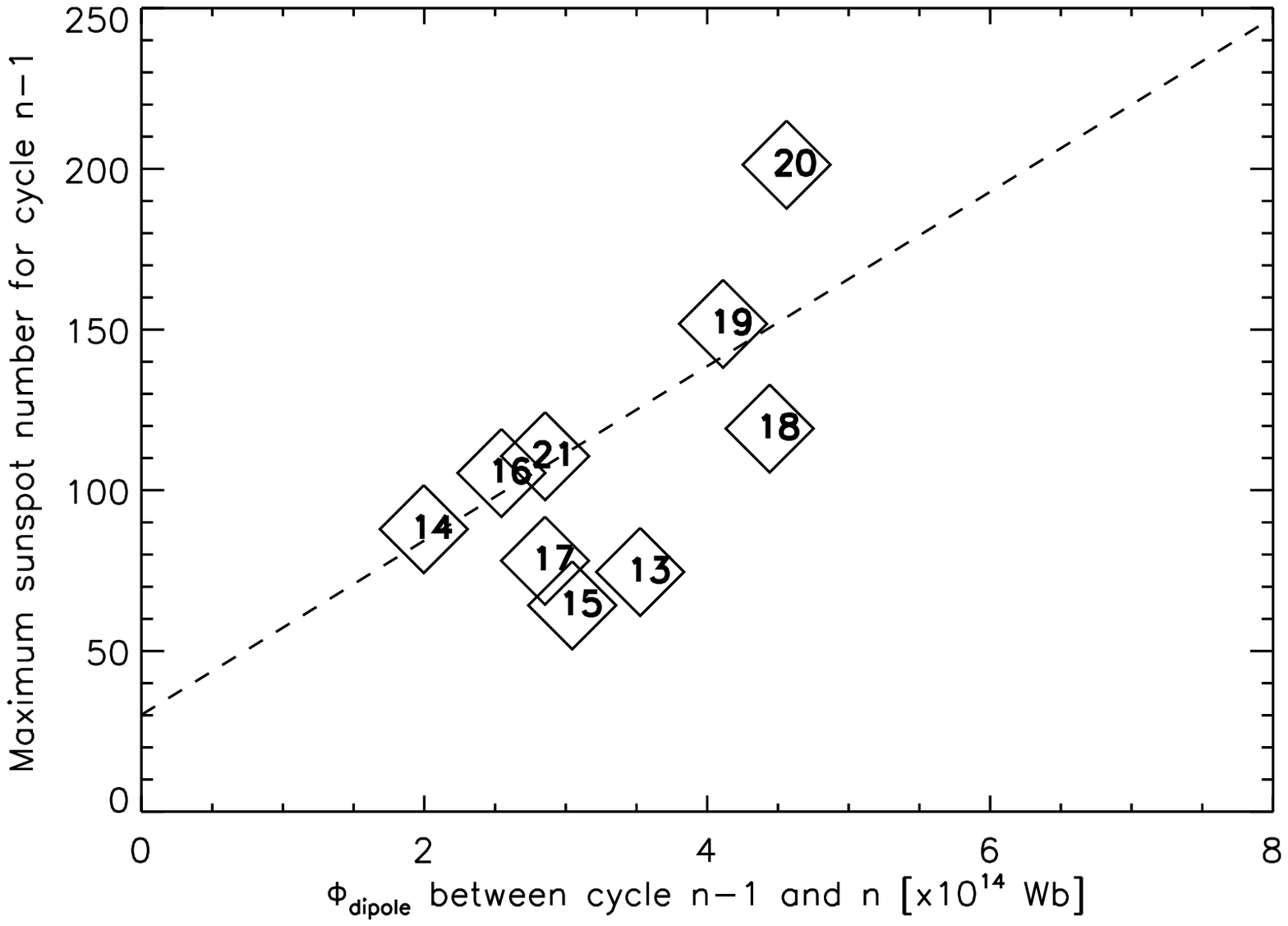}
   \includegraphics[scale=0.5]{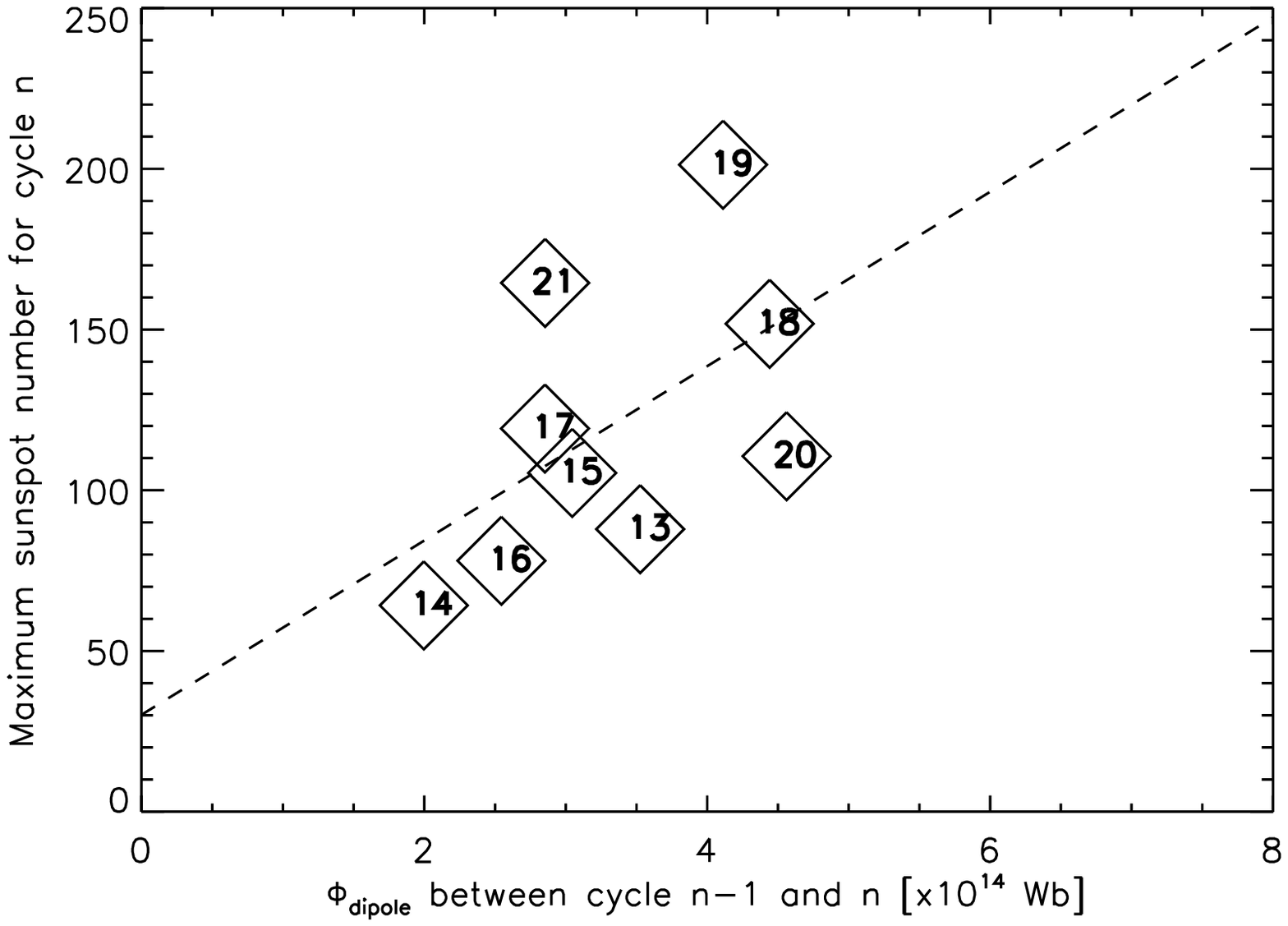}}\\
   {\includegraphics[scale=0.5]{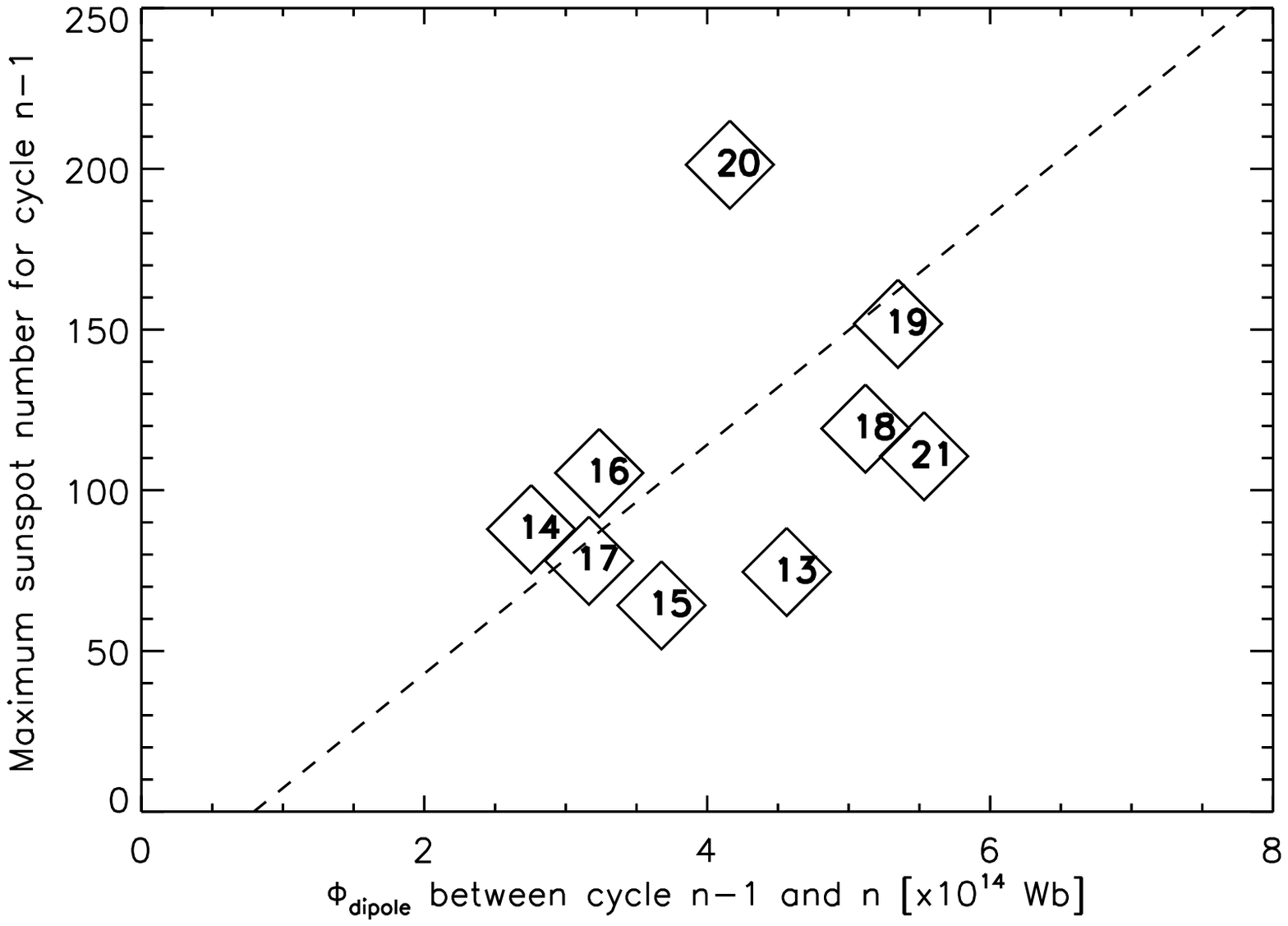}
   \includegraphics[scale=0.5]{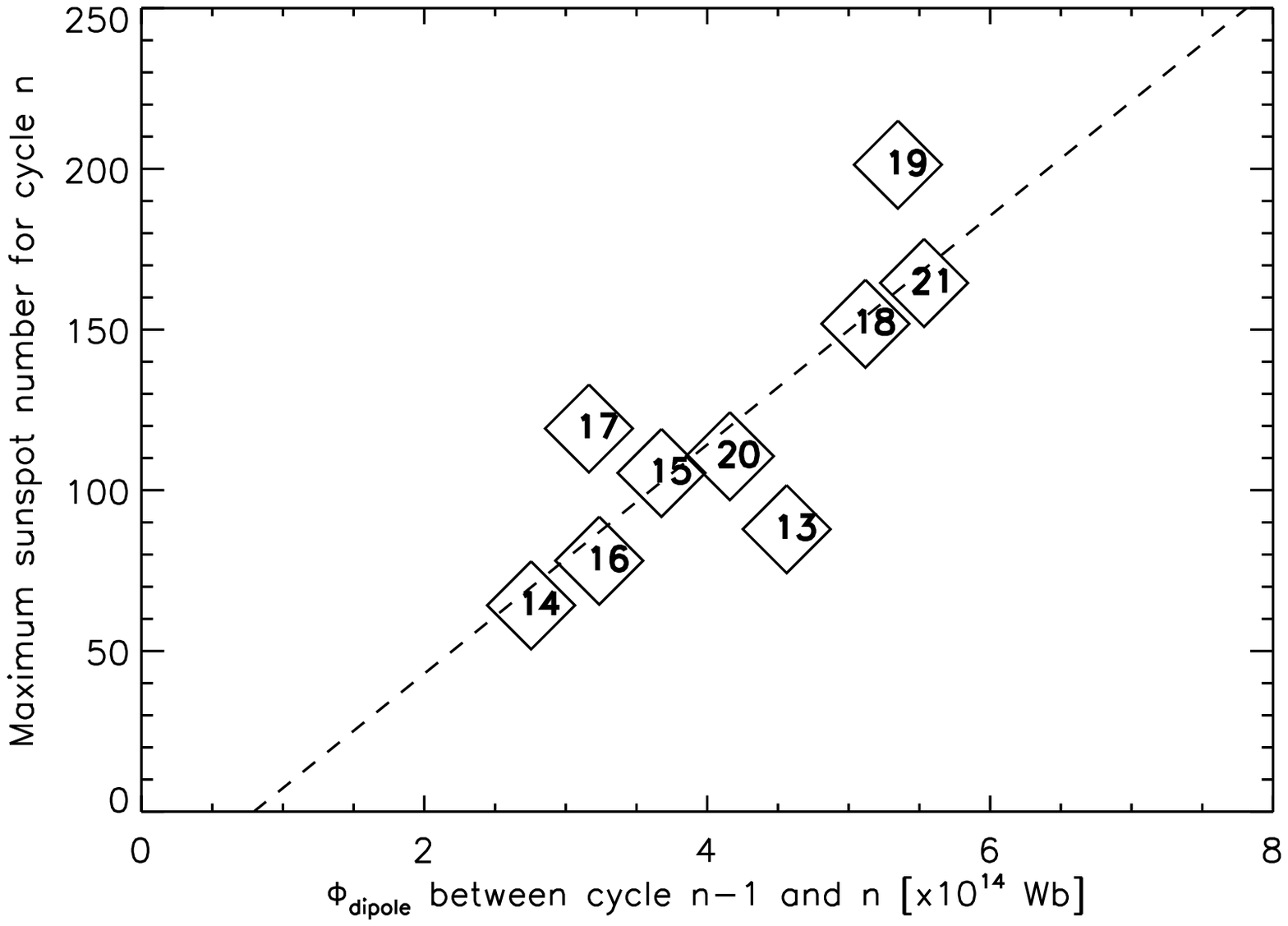}}
      \caption{Correlation diagrams for the maxima of \phidip\ during
              activity minima between cycles $n-1$ and $n$ from the SFT
              simulation vs. the maxima of the sunspot number of the the
              cycle preceding the minimum ($n-1$, {\em left panels}) and
              of the subsequent cycle ($n$, {\em right panels}),
              respectively.  {\em Upper row}: case without inflows; {\em
              lower row}: reference case with inflows. Numbers within
              symbols give the index, $n$, of the cycle following the
              respective activity minimum. Linear regressions are
              indicated by the dashed lines.}
         \label{fig:cc_dipole_sunspots}
\end{figure*}

\subsection{Relation between axial dipole and cycle strength}
\label{subsec:cycle_strength}

There is an empirical correlation between quantities related to the
polar fields around activity minimum and the amplitude of the subsequent
activity cycle.  For instance, considering the values of the open
heliospheric flux near activity minimum as inferred from geomagnetic
variations \citep{Lockwood:etal:2009} and the maximum sunspot number
(based on smoothed monthly mean values%
\footnote{\tt ftp://ftp.ngdc.noaa.gov/STP/SOLAR\_DATA/SUNSPOT\_NUMBERS/docs/maxmin.new}%
) of the {\em subsequent} cycle, \rmax, we find $r=0.87$ ($p=0.0025$)
for cycles 13--21. This is consistent with the results of
\citet{Wang:Sheeley:2009}.
On the other hand, the correlation with the maximum
sunspot number of the {\em preceding} cycle is insignificant $(r=0.55,\,
p=0.12)$. These results indicate that the polar fields reflect the
poloidal source of the toroidal field of the next cycle, which is
consistent with the interpretation of the solar cycle in terms of a
BL-type dynamo.

In order to evaluate the effect of the latitudinal inflows on this
correlation and thus infer their viability as a nonlinear feedback
mechanism for BL-type dynamos, we consider the correlations between the
maxima of \phidip\ as provided by SFT simulations and the values of
\rmax\ of the subsequent and preceding cycles, respectively.
Fig.~\ref{fig:cc_dipole_sunspots} shows these relations for the
simulations with and without inflows, respectively. In latter case, the
correlation is moderate for \rmax\ of the preceding cycle (upper left
panel; $r=0.68,\;p=0.044$) and insignificant for the subsequent cycle
(upper right panel; $r=0.54,\;p=0.13$), which is in striking contrast to
the empirical correlations. Including the inflows completely changes the
picture: while the correlation of \phidip\ with \rmax\ of the preceding
cycle becomes insignificant (lower left panel; $r=0.38,\;p=0.31$), the
correlation with \rmax\ of the subsequent cycle is highly significant in
this case (lower right panel; $r=0.82,\;p=0.0065$), which is comparable
to the empirical correlation based on the open heliospheric flux.

These results show that the inflows towards the activity belts
provide an important nonlinearity in the evolution of the Sun's poloidal
field at the solar surface. This conclusion follows essentially from the
magnitude of the modeled inflows and is largely independent of the model
parameters. Furthermore, our results strongly suggest that the
latitudinal inflows towards the activity belts have a dominant influence
on the strength of the subsequent cycle and provide a nonlinear feedback
mechanism for a BL-type dynamo.  While other nonlinear processes might
also play a role, the high correlations found here suggest that their
effect is probably rather limited. Further work is going to assess
the extent to which the good correlation with the observed cycle
strengths found here depends on the SFT parameters and on our model of
the inflows. Here we simply note that the parameters used in this study
are essentially those of \citet[][SFT parameters]{Cameron10} and
of \citet[][inflow model]{Cameron10b}, which all are based on
observational constraints.
 
The high sensitivity of the axial dipole strength on the amplitude of
the inflows is consistent with the large variation of the cycle
amplitudes in the historical record, indicating also that rather weak
fluctuations of the surface distribution of active regions could
temporarily switch off the BL-type dynamo and may drive the system into
an extended minimum state.

\section{Conclusion}

Our SFT simulations show that magnetic-field-dependent latitudinal
inflows converging towards the activity belts significantly affect the
build-up of the polar field by modifying the cross-equator transport of
magnetic flux. The resulting amplitudes of the magnetic flux contained
in the axial dipole component during activity minima correlate well with
the empirically derived values of the open heliospheric flux during
these periods. The inflows strengthen the axial dipole in weaker cycles,
for which parts of the inflows provide enhanced cross-equator transport
of magnetic flux. For strong cycles, the reduction of the tilt angles of
bipolar magnetic regions by the converging inflows dominates and leads to
a weakening of the axial dipole.  Consistent with the empirical results,
the SFT simulations including latitudinal inflows show a strong
correlation between the axial dipole around activity minimum and the
observed maximum sunspot number of the subsequent cycle. This indicates
that the inflows are a key ingredient in determining the amplitude of
solar cycles by providing a nonlinear feedback mechanism for the
saturation of a Babcock-Leighton-type dynamo mechanism.

 \bibliographystyle{aa.bst}
 \bibliography{aa19914-12.bbl}


\end{document}